\documentclass[a4paper,11pt]{article}
\usepackage{pos}
\usepackage{comment}
\usepackage{lineno}

\title{Detection of the 2021 Outburst of RS Ophiuchi with the LST-1}

\author*[a]{Yukiho Kobayashi}
\author[b]{Arnau Aguasca-Cabot}
\author[c]{Mar\'ia Isabel Bernardos Martín}
\author[d]{David Green}
\author[c]{Rubén López-Coto}

\affiliation[a]{Institute for Cosmic Ray Research, The University of Tokyo, 5-1-5, Kashiwa-no-ha, Kashiwa, Chiba 277-8582, Japan}
\affiliation[b]{Departament de Física Quàntica i Astrofísica, Institut de Ciències del Cosmos, Universitat de Barcelona, IEEC-UB, Martí i Franquès, 1, 08028, Barcelona, Spain}
\affiliation[c]{Instituto de Astrofísica de Andalucía-CSIC, Glorieta de la Astronomía s/n, 18008, Granada, Spain}
\affiliation[d]{Max-Planck-Institut für Physik, Föhringer Ring 6, 80805 München, Germany}

\onbehalf{for the CTA-LST project} 

\emailAdd{yukihok@icrr.u-tokyo.ac.jp}

\abstract{Novae are luminous explosions in close binaries which host a white dwarf and a companion donor star. They are triggered by a thermonuclear runaway when the white dwarf accretes a critical amount of matter from the secondary. Though novae are established as high-energy gamma-ray emitters through observations by the Fermi Large Area Telescope (LAT), the origin of the gamma-ray emission, whether it is hadronic or leptonic, had been under intense debate until very recently. RS Ophiuchi (RS Oph) is a well-known recurrent symbiotic nova with a recurrence time scale of 15 years. The most recent outburst of RS Oph in 2021 brought the first detection of very-high-energy (VHE) gamma rays from a nova ever. The first Large-Sized Telescope prototype (LST-1) of the Cherenkov Telescope Array observed this historic event along with H.E.S.S. and MAGIC.
The LST-1 observations in the first days after the burst onset show a clear VHE gamma-ray signal from RS Oph.
The low energy threshold of LST-1 allows us to reconstruct the RS Oph gamma-ray spectrum down to $\sim$30 GeV, providing the best connection of the VHE gamma-ray data to the Fermi LAT energy range. The results from the analysis of the LST-1 observations are consistent
with those obtained with H.E.S.S. and MAGIC, and also support a hadronic origin for the observed gamma-ray fluxes.
In this contribution, we will present the analysis results of the LST-1 observations of the 2021 outburst of RS Oph.}

\ConferenceLogo{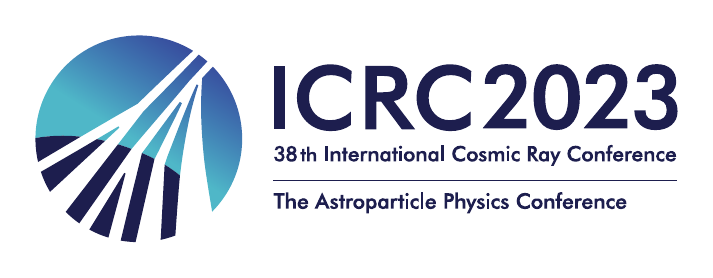}

\FullConference{%
38th International Cosmic Ray Conference (ICRC2023)\\
  26 July - 3 August, 2023\\
  Nagoya, Japan}

\begin{document}
\maketitle


\section{Introduction}
Novae are luminous eruptions observed in close binaries where a white dwarf (WD) is interacting with its stellar companion.
The explosion is powered by a thermonuclear runaway ignited on the surface of the WD when it accretes a critical amount of matter from the companion donor star.
Novae have been revealed to be high-energy gamma-ray emitters through observations by the Fermi Large Area Telescope (LAT), starting with the detection of gamma-ray emission from V407 Cyg in 2010 \cite{fermi-novae}.
Though more than a dozen novae have been detected by the Fermi LAT, the mechanism responsible for the production of gamma-ray emission in novae had remained unclear until very recently.
In August 2021, RS Ophiuchi (RS Oph), a well-known recurrent nova, erupted after an interval of 15 years since the previous outburst in 2006.
Noteworthy, the 2021 outburst of RS Oph was detected in the very-high-energy (VHE, >100~GeV) regime by Imaging Atmospheric Cherenkov Telescopes (IACTs) such as High Energy Stereoscopic System (H.E.S.S.) and the Major Atmospheric Gamma Imaging Cherenkov (MAGIC) telescopes, which marked the event as the first nova explosion that was detected in VHE gamma rays in history \cite{hess2022time, acciari2022proton}.

The Cherenkov Telescope Array (CTA) is the next-generation IACT array consisting of three kinds of telescopes with different sizes. 
Among them, the Large-Sized Telescopes (LSTs), equipped with a large mirror dish of 23~m in diameter and a sensitive focal-plane camera with fast readout, dominate the lower energy band of CTA down to $\sim$20 GeV.
The LST prototype (LST-1) for CTA, built on the CTA northern site, La Palma in the Canary Islands of Spain, has been in its commissioning phase and performing astrophysical observations since 2019 \cite{2022icrc.confE.872C}.
The LST-1 observed the historic nova outburst of RS Oph in 2021.
In this contribution, we present the analysis results of the LST-1 observations.

\section{Observations}
RS Oph is a well-known recurrent nova with a relatively short recurrence time-scale of $\sim$15 years.
RS Oph is also known to be a symbiotic system, where the WD is embedded in the outflow from the red giant companion.
Thanks to the short recurrence time-scale, outbursts of RS Oph have been recorded several times in history \cite{2010ApJS..187..275S}.
The outburst in 2006 was especially well studied at various wavelengths, such as optical, radio, and x-ray, but it was before the launch of Fermi \cite{tatischeff2007evidence}.

On August 8th 2021, RS Oph was reported to be in a new burst state from observations in the optical and gamma-ray bands \cite{aavso_alert, vsnet_alert, 2021ATel14834, cheung2022fermi}.
In response to these alerts, the LST-1 started observations of RS Oph on August 9th, just a day after the burst onset.
The observations were performed under proper conditions, i.e., a clear and dark sky, until August 12th.
These good-quality data in the first nights amount to 6.4 hours of effective observational time.
From August 13th, however, bad weather and enhanced moonlight conditions prevented further data taking.
We resumed observations on August 29th and continued until September 2nd under good observational conditions, amounting to about four more hours.
The zenith angle range of the observations was between 35$^\circ$ and 64$^\circ$.
The observations were performed in the so-called \textit{wobble} mode with a wobble offset of 0.4$^\circ$ \cite{fomin1994new}.
The LST-1 observations of the 2021 outburst of RS Oph are summarized in Table~\ref{obs_summary}.

\begin{table}[htbp]
 \caption{Summary of the LST-1 observations of RS Oph during its 2021 outburst}
 \centering
  \begin{tabular}{ccccc}
   \hline
   date & observation & zenith range \\
   & time [h] & [deg]  \\
   \hline \hline
   9 Aug. 2021 & 1.4 h & 35-42 \\
   10 Aug. 2021 & 2.7 h & 35-59\\
   12 Aug. 2021 & 2.3 h & 35-55 \\
   13 Aug. 2021 & 1.3 h & 36-54 \\
   14 Aug. 2021 & 1.5 h & 35-46 \\
   15 Aug. 2021 & 1.3 h & 41-57 \\
   29 Aug. 2021 & 1.0 h & 46-58\\
   30 Aug. 2021 & 1.5 h & 40-57\\
   1 Sep. 2021 & 0.3 h & 56-64 \\
   2 Sep. 2021 & 1.3 h & 41-57 \\
   \hline
  \end{tabular}
  \label{obs_summary}
\end{table}

\section{Analysis}

The LST-1 observations are reduced following the standard LST analysis procedure \cite{2023arXiv230612960P}.
The analysis is performed with \texttt{cta-lstchain}\footnote{https://github.com/cta-observatory/cta-lstchain}, a dedicated LST analysis software based on the CTA low-level analysis pipeline \texttt{ctapipe} \cite{ruben_lopez_coto_2022_6344674, lst_performance_icrc2021, karl_kosack_2021_5720333, ctapipe-icrc-2021}.
The \texttt{cta-lstchain} performs all the analysis steps from calibration and signal extraction to the reconstruction of events.
For the reconstruction of the energy and direction of primary particles, random forest (RF) algorithms are adopted.
An RF is also used to reject background cosmic-ray events by giving a score called \textit{gammaness} to each event, which represents how likely the primary particle is a gamma ray.
The so-called source-independent approach, a method to reconstruct events without an assumption on the source position, is applied in this work.
The LST-1 data are reduced to the DL3 format, which is subsequently fed to \texttt{gammapy}\footnote{https://gammapy.org}, the official science analysis tool for CTA, for computing the gamma-ray spectrum of the source and the source flux light-curve \cite{acero_fabio_2022_7311399, gammapy:2017}.

\begin{figure}[htbp]
    \centering
    \includegraphics[width=.55\linewidth]{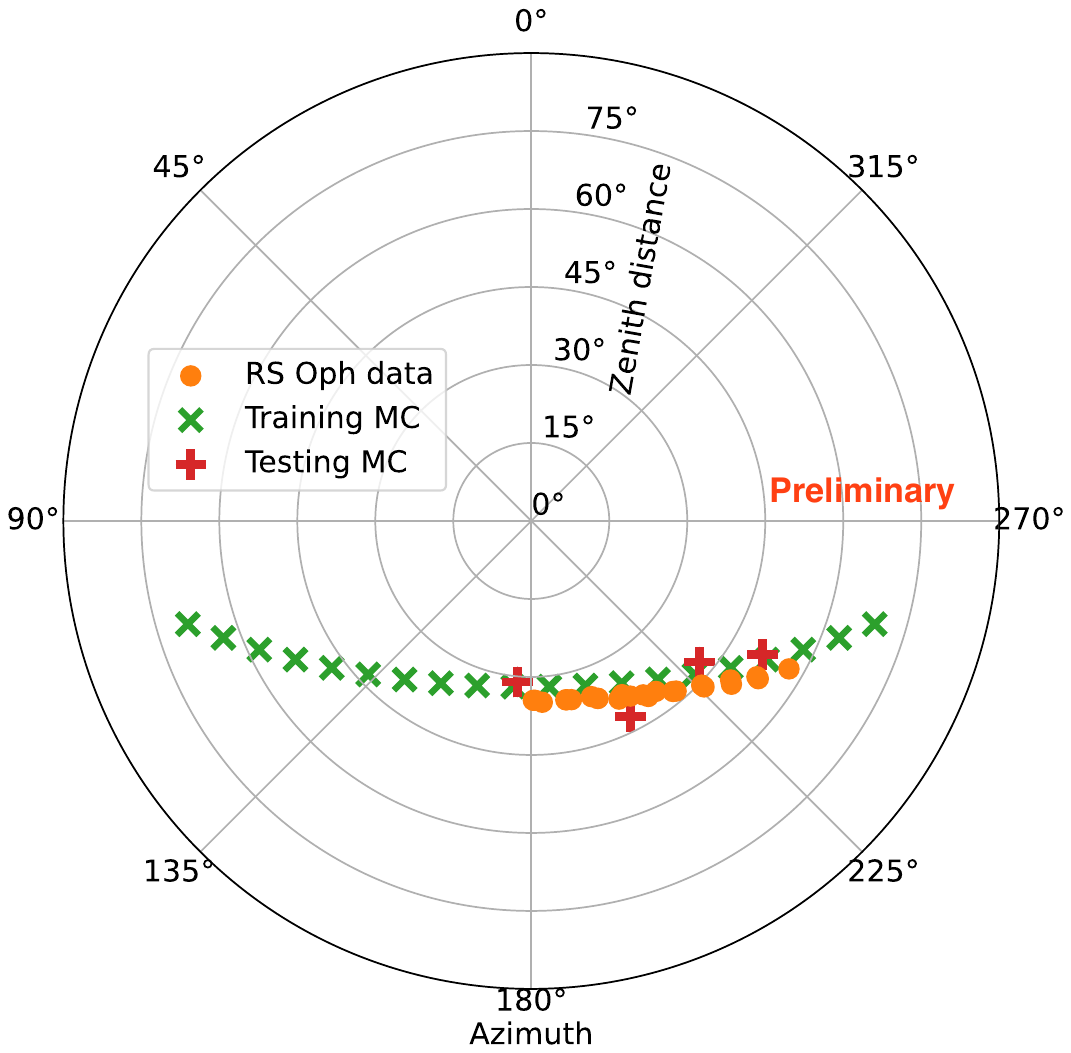}
    \caption{The pointing directions of the data taken on 9th, 10th, and 12th of August 2021, the closest testing MC nodes to the data and the training MC nodes on the plane of the zenith angle and the azimuth angle.}
    \label{fig:MC-nodes}
\end{figure}

Monte Carlo (MC) simulations are prepared to train the RF algorithms and evaluate the telescope's instrumental response functions (IRFs).
The MC simulations that are used in this work are generated according to a standard procedure for the LST analysis, but are tuned to the RS Oph observations \cite{2023arXiv230612960P}.
For instance, the amount of night sky background contamination in the simulations is adjusted to the observations at the camera image level.
Simulations are produced for different positions in the sky to take into account the dependence of the telescope performance for different pointing directions.
As shown in Figure~\ref{fig:MC-nodes}, the simulations to train the RF algorithms are generated along a declination path close to that of RS Oph and those for testing the telescope performance are produced in a grid of $\cos{\rm ZD}$ and $\sin{\delta}$, where $\rm ZD$ is the zenith angle and $\delta$ is an angle between the geomagnetic field and the pointing direction.
The IRFs of the LST-1 are evaluated in each testing MC node and the closest node to each observational run is used to compute gamma-ray flux.
For processing MC simulations, \texttt{lstmcpipe}, a dedicated package for reduction of LST MC simulations, is adopted \cite{vuillaume_thomas_2022_7180216, garcia2022lstmcpipe}.

For the spectral analysis, the energy-dependent cuts on \textit{gammaness} are defined from the MC simulations so that gamma rays are kept with an efficiency of 50\%.
Directional cuts are also applied energy-dependently with a 70\% gamma-ray efficiency.
The energy threshold of the LST-1 observations for this work is evaluated from the true energy distribution of the simulated gamma-ray events surviving all the event selection criteria.
Though the energy threshold depends on the zenith angle, an average value over the LST-1 observations is found to be $\sim$30~GeV, assuming that RS Oph has a power-law spectrum with an index of $\sim$4, as indicated from the observations by H.E.S.S. and MAGIC.
This is the lowest energy threshold for the observations of VHE gamma rays from RS Oph among the IACTs, and thus the LST-1 provides the best connection of the VHE gamma-ray data to the Fermi LAT energy range.
The gamma-ray flux of RS Oph is reconstructed assuming a point-like source with a power-law spectral model, $\Phi(E) = \Phi_0 \left({E}/{E_{\rm ref}}\right)^{-p}$, where the spectrum is normalized at $\Phi(E_{\rm ref}) = \Phi_0$ with a reference energy $E_{\rm ref} = 130$~GeV.


\section{Results}
\begin{figure}[htbp]
    \begin{minipage}[c]{.5\linewidth}
        \centering
        \includegraphics[width=1.\linewidth]{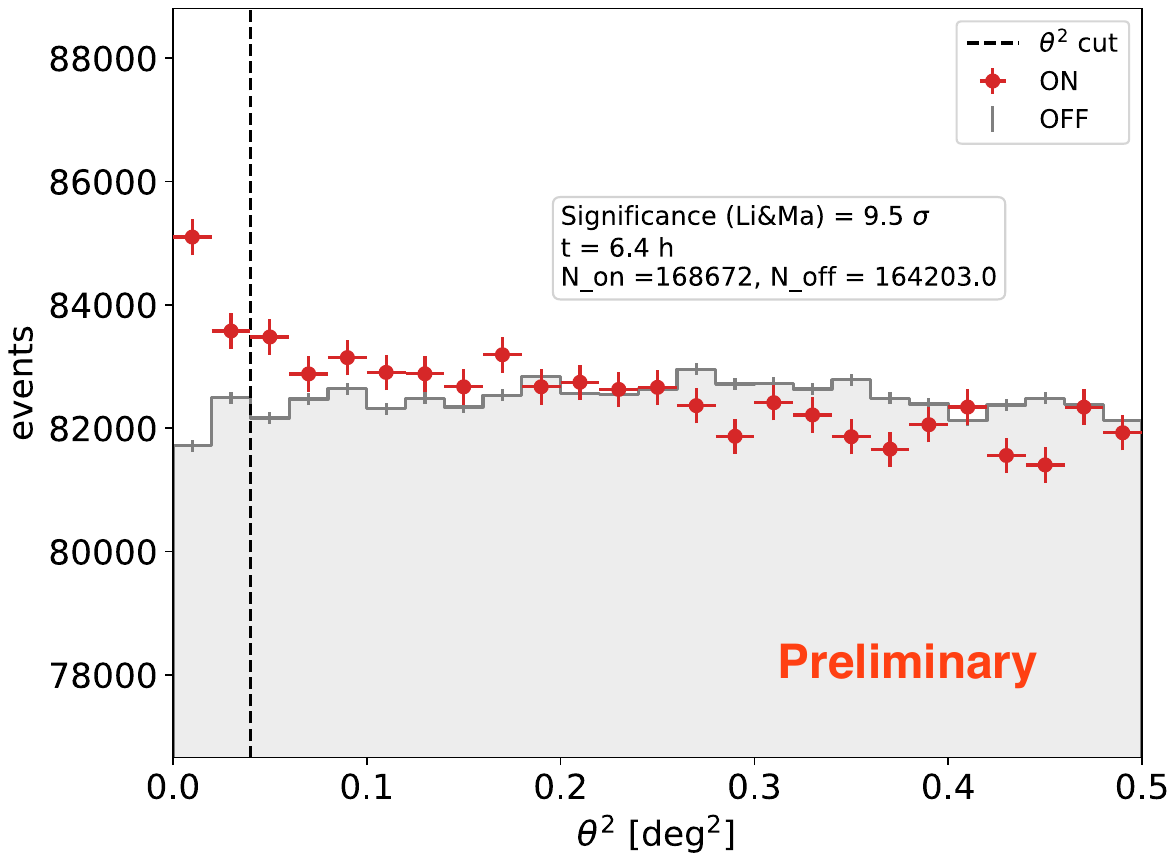}
    \end{minipage}
    \begin{minipage}[c]{.5\linewidth}
        \centering
        \includegraphics[width=1.\linewidth]{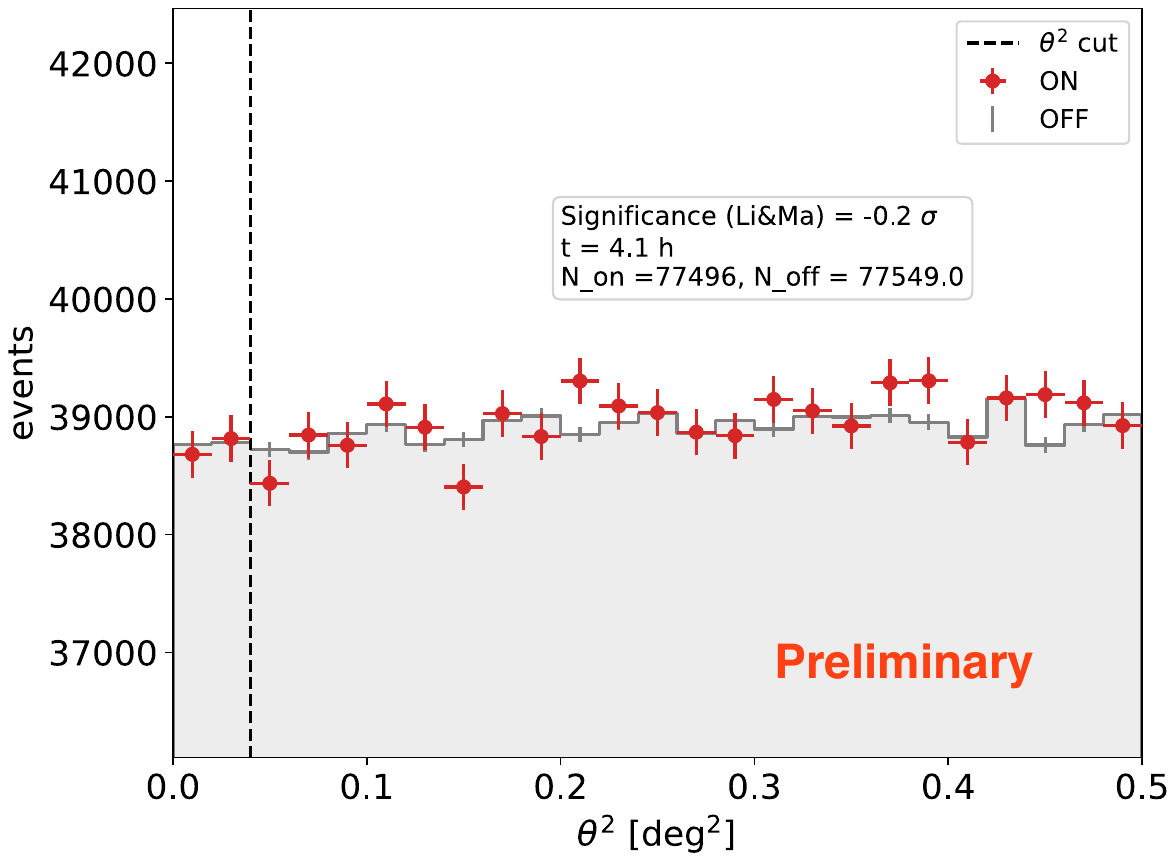}
    \end{minipage}
    \caption{Distribution of events recorded for RS Oph as a function of $\theta^2$, the squared angular distance between the reconstructed arrival direction of the gamma-ray candidate events and the position of RS Oph.
    Events with $\textit{gammaness} > 0.6$ and reconstructed energies between 30~GeV and 1~TeV are shown.
    \textit{Left}: Observations between August 9th and August 12th of 2021. \textit{Right}: Observations between August 29th and September 2nd.}
    \label{fig:theta2}
\end{figure}

\begin{figure}[htbp]
    \centering
    \includegraphics[width=0.85\linewidth]{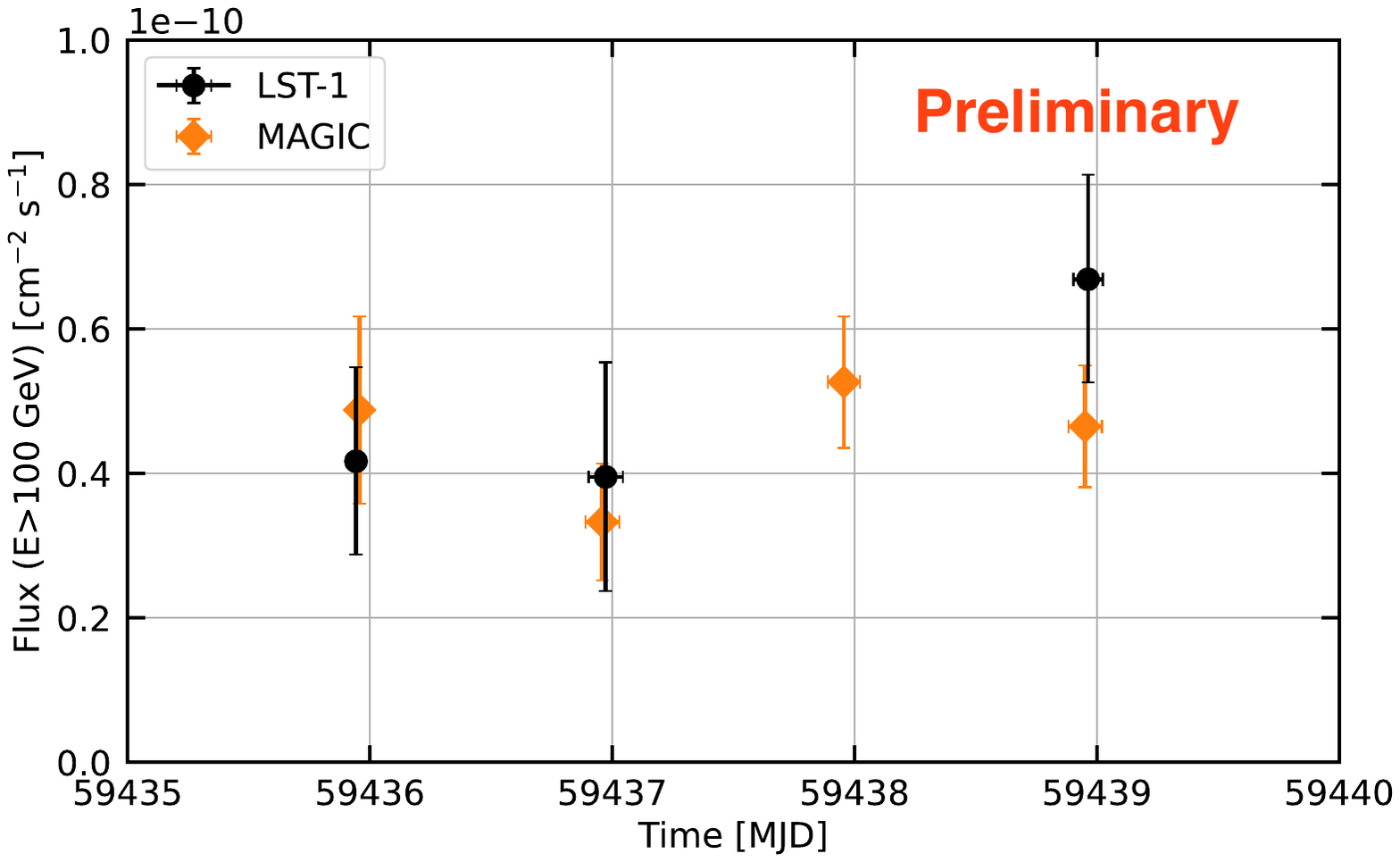}
    \caption{Daily gamma-ray flux of RS Oph during the first nights of its 2021 outburst reconstructed with the LST-1 in comparison with the MAGIC results.}
    \label{fig:lc}
\end{figure}

Figure~\ref{fig:theta2} shows the distribution of the squared angular distance between the reconstructed arrival direction of the gamma-ray candidate events obtained from the LST-1 observations and the position of RS Oph.
Events with $\textit{gammaness} > 0.6$ and reconstructed energies between 30~GeV and 1~TeV are shown.
From the observations between August 9th and August 12th, a gamma-ray signal from the direction of RS Oph is clearly detected at a statistical significance of 9.5 $\sigma$.
The signal is also visible each night at a significance above 5 $\sigma$.
From the observations on and after 29th August, on the other hand, no significant excess is found.
The light curve during the first nights after the burst onset is computed at energies above 100~GeV and shown in Figure~\ref{fig:lc}.
The light curve that is reconstructed from the LST-1 observations is compatible with a constant flux during the first nights and it is in good agreement with the MAGIC results.


\section{Conclusion}
The LST-1 performed observations of RS Oph during its 2021 outburst and acquired good-quality data during the first nights after the eruption.
The LST-1 data are analyzed with the standard LST analysis tools and dedicated all-sky MC simulations tuned to the RS Oph observations.
The gamma-ray emission from RS Oph is firmly detected by the LST-1 during the first nights after the burst onset at a statistical significance of 9.5 $\sigma$. 
Noteworthy, the LST-1 achieves the lowest energy threshold for the VHE gamma-ray observations of the 2021 outburst of RS Oph among the IACTs $\sim$30~GeV, which gives the best connection of the VHE gamma-ray data to the Fermi LAT energy band.
A detailed interpretation of the LST-1 observations with dedicated modeling is in progress and the results will be presented in a forthcoming publication.

\newpage
\begin{center}

\bigskip
\bigskip

{\bf LST Acknowledgements }

\bigskip

We gratefully acknowledge financial support from the following agencies and organisations:

\bigskip


\small
Ministry of Education, Youth and Sports, MEYS  LM2015046, LM2018105, LTT17006, EU/MEYS CZ.02.1.01/0.0/0.0/16\_013/0001403, CZ.02.1.01/0.0/0.0/18\_046/0016007 and CZ.02.1.01/0.0/0.0/16\_019/0000754, Czech Republic; 
Max Planck Society, German Bundesministerium f{\"u}r Bildung und Forschung (Verbundforschung / ErUM), Deutsche Forschungsgemeinschaft (SFBs 876 and 1491), Germany;
Istituto Nazionale di Astrofisica (INAF), Istituto Nazionale di Fisica Nucleare (INFN), Italian Ministry for University and Research (MUR);
ICRR, University of Tokyo, JSPS, MEXT, Japan;
JST SPRING - JPMJSP2108;
Narodowe Centrum Nauki, grant number 2019/34/E/ST9/00224, Poland;
The Spanish groups acknowledge the Spanish Ministry of Science and Innovation and the Spanish Research State Agency (AEI) through the government budget lines PGE2021/28.06.000X.411.01, PGE2022/28.06.000X.411.01 and PGE2022/28.06.000X.711.04, and grants PGC2018-095512-B-I00, PID2019-104114RB-C31, PID2019-107847RB-C44, PID2019-104114RB-C32, PID2019-105510GB-C31, PID2019-104114RB-C33, PID2019-107847RB-C41, PID2019-107847RB-C43, PID2019-107988GB-C22;
the “Centro de Excelencia Severo Ochoa” program through grants no. CEX2021-001131-S, CEX2019-000920-S;
the “Unidad de Excelencia Mar{\'i}a de Maeztu” program through grants no. CEX2019-000918-M, CEX2020-001058-M;
the “Juan de la Cierva-Incorporaci{\'o}n” program through grant no. IJC2019-040315-I. They also acknowledge the “Programa Operativo” FEDER 2014-2020, Consejer{\'i}a de Econom{\'i}a y Conocimiento de la Junta de Andaluc{\'i}a (Ref. 1257737), PAIDI 2020 (Ref. P18-FR-1580) and Universidad de Ja{\'e}n;
“Programa Operativo de Crecimiento Inteligente” FEDER 2014-2020 (Ref.~ESFRI-2017-IAC-12), Ministerio de Ciencia e Innovaci{\'o}n, 15\% co-financed by Consejer{\'i}a de Econom{\'i}a, Industria, Comercio y Conocimiento del Gobierno de Canarias;
the “CERCA” program of the Generalitat de Catalunya;
and the European Union’s “Horizon 2020” GA:824064 and NextGenerationEU;
We acknowledge the Ramon y Cajal program through grant RYC-2020-028639-I and RYC-2017-22665;
State Secretariat for Education, Research and Innovation (SERI) and Swiss National Science Foundation (SNSF), Switzerland;
The research leading to these results has received funding from the European Union's Seventh Framework Programme (FP7/2007-2013) under grant agreements No~262053 and No~317446.
This project is receiving funding from the European Union's Horizon 2020 research and innovation programs under agreement No~676134.
ESCAPE - The European Science Cluster of Astronomy \& Particle Physics ESFRI Research Infrastructures has received funding from the European Union’s Horizon 2020 research and innovation programme under Grant Agreement no. 824064.

\end{center}




%
%

\newpage
\bibliographystyle{JHEP}
\small
\bibliography{bibliography}

\providecommand{\href}[2]{#2}\begingroup\raggedright\begin{thebibliography}{10}

\bibitem{fermi-novae}
{The Fermi-LAT Collaboration}, \emph{Fermi establishes classical novae as a
  distinct class of gamma-ray sources}, {\emph{Science} {\bfseries 345} (2014)
  554}.

\bibitem{hess2022time}
{H.E.S.S. Collaboration}, \emph{Time-resolved hadronic particle acceleration in
  the recurrent nova rs ophiuchi}, {\emph{Science} {\bfseries 376} (2022) 77}.

\bibitem{acciari2022proton}
{MAGIC Collaboration}, \emph{Proton acceleration in thermonuclear nova
  explosions revealed by gamma rays}, {\emph{Nature Astronomy} {\bfseries 6}
  (2022) 689}.

\bibitem{2022icrc.confE.872C}
{CTA-LST Project}, \emph{{Status and results of the prototype LST of CTA}},  in
  \emph{37th International Cosmic Ray Conference}, p.~872, Mar., 2022.

\bibitem{2010ApJS..187..275S}
B.E.~Schaefer, \emph{Comprehensive photometric histories of all known galactic
  recurrent novae}, {\emph{The Astrophysical Journal Supplement Series}
  {\bfseries 187} (2010) 275}.

\bibitem{tatischeff2007evidence}
V.~Tatischeff and M.~Hernanz, \emph{Evidence for nonlinear diffusive shock
  acceleration of cosmic rays in the 2006 outburst of the recurrent nova rs
  ophiuchi}, {\emph{The Astrophysical Journal} {\bfseries 663} (2007) L101}.

\bibitem{aavso_alert}
S.~Beck, ``Aavso alert notice 752: Rare outburst of recurrent nova rs
  ophiuchi.'' \url{https://www.aavso.org/aavso-alert-notice-752}.

\bibitem{vsnet_alert}
K.~Geary, ``Vsnet-alert notice, 26131.''
  \url{http://ooruri.kusastro.kyoto-u.ac.jp/mailarchive/vsnet-alert/26131}.

\bibitem{2021ATel14834}
C.C.~{Cheung}, S.~{Ciprini} and T.J.~{Johnson}, \emph{{Fermi-LAT Gamma-ray
  Detection of the Recurrent Nova RS Oph}}, {\emph{The Astronomer's Telegram}
  {\bfseries 14834} (2021) 1}.

\bibitem{cheung2022fermi}
C.~Cheung, T.~Johnson, P.~Jean, M.~Kerr, K.~Page, J.~Osborne et~al.,
  \emph{Fermi lat gamma-ray detection of the recurrent nova rs ophiuchi during
  its 2021 outburst}, {\emph{The Astrophysical Journal} {\bfseries 935} (2022)
  44}.

\bibitem{fomin1994new}
V.~Fomin et~al., \emph{New methods of atmospheric cherenkov imaging for
  gamma-ray astronomy. i. the false source method}, {\emph{Astroparticle
  Physics} {\bfseries 2} (1994) 137}.

\bibitem{2023arXiv230612960P}
{CTA-LST Project}, \emph{{Observations of the Crab Nebula and Pulsar with the
  Large-Sized Telescope Prototype of the Cherenkov Telescope Array}},
  {\emph{arXiv e-prints} (2023) arXiv:2306.12960}.

\bibitem{ruben_lopez_coto_2022_6344674}
R.~Lopez-Coto et~al., \emph{cta-observatory/cta-lstchain: v0.9.4},  Mar., 2022.
\newblock 10.5281/zenodo.6344674.

\bibitem{lst_performance_icrc2021}
R.~López-Coto et~al., \emph{{Physics Performance of the Large Size Telescope
  prototype of the Cherenkov Telescope Array}},  in \emph{Proceedings, 37th
  International Cosmic Ray Conference}, vol.~395, p.~806, 2021.

\bibitem{karl_kosack_2021_5720333}
K.~Kosack et~al., \emph{cta-observatory/ctapipe: v0.12.0},  Nov., 2021.
\newblock 10.5281/zenodo.5720333.

\bibitem{ctapipe-icrc-2021}
M.~Nöthe, K.~Kosack, L.~Nickel and M.~Peresano, \emph{Prototype open event
  reconstruction pipeline for the cherenkov telescope array},  in
  \emph{Proceedings, 37th International Cosmic Ray Conference}, vol.~395, 2021.

\bibitem{acero_fabio_2022_7311399}
F.~Acero et~al., \emph{Gammapy: Python toolbox for gamma-ray astronomy},  Nov.,
  2022.
\newblock 10.5281/zenodo.7311399.

\bibitem{gammapy:2017}
C.~{Deil} et~al., \emph{{Gammapy - A prototype for the CTA science tools}},  in
  \emph{35th International Cosmic Ray Conference (ICRC2017)}, vol.~301 of
  \emph{International Cosmic Ray Conference}, p.~766, Jan., 2017.

\bibitem{vuillaume_thomas_2022_7180216}
T.~Vuillaume, E.~Garcia and L.~Nickel, \emph{lstmcpipe},  Oct., 2022.
\newblock 10.5281/zenodo.7180216.

\bibitem{garcia2022lstmcpipe}
E.~Garcia, T.~Vuillaume and L.~Nickel, \emph{The lstmcpipe library},  2022.

\end{thebibliography}\endgroup

\clearpage
\section*{Full Author List: CTA-LST Project}

\tiny{\noindent
K. Abe$^{1}$,
S. Abe$^{2}$,
A. Aguasca-Cabot$^{3}$,
I. Agudo$^{4}$,
N. Alvarez Crespo$^{5}$,
L. A. Antonelli$^{6}$,
C. Aramo$^{7}$,
A. Arbet-Engels$^{8}$,
C.  Arcaro$^{9}$,
M.  Artero$^{10}$
% ...
}\\

\tiny{\noindent
$^{1}$Department of Physics, Tokai University.
$^{2}$Institute for Cosmic Ray Research, University of Tokyo.
$^{3}$Departament de Física Quàntica i Astrofísica, Institut de Ciències del Cosmos, Universitat de Barcelona, IEEC-UB.
$^{4}$Instituto de Astrofísica de Andalucía-CSIC.
$^{5}$EMFTEL department and IPARCOS, Universidad Complutense de Madrid.
$^{6}$INAF - Osservatorio Astronomico di Roma.
$^{7}$INFN Sezione di Napoli.
$^{8}$Max-Planck-Institut für Physik.
$^{9}$INFN Sezione di Padova and Università degli Studi di Padova.
$^{10}$Institut de Fisica d'Altes Energies (IFAE), The Barcelona Institute of Science and Technology.
% ...
}
\end{document}